\documentstyle{article}
\input{epsf}
\newcommand{\be}{\begin{equation}}
\newcommand{\ee}{\end{equation}}
\newcommand{\bea}{\begin{eqnarray}}
\newcommand{\eea}{\end{eqnarray}}
\newcommand{\ba}{\begin{array}}
\newcommand{\ea}{\end{array}}

\def\bbox{{\,\lower0.9pt\vbox{\hrule \hbox{\vrule height 0.2 cm
\hskip 0.2 cm \vrule height 0.2 cm}\hrule}\,}}
\newcommand{\dsl}{\pa \kern-0.5em /}

\newcommand{\nn}{\nonumber \\}

\newcommand{\EQ}{\begin{equation}}
\newcommand{\EN}{\end{equation}}

\def\bbox{{\,\lower0.9pt\vbox{\hrule \hbox{\vrule height 0.2 cm
\hskip 0.2 cm \vrule height 0.2 cm}\hrule}\,}}

\newcommand{\pa}{\partial}

\def\today{\ifcase\month\or
  January\or February\or March\or April\or May\or June\or
  July\or August\or September\or October\or November\or December\fi
 \space\number\day, \number\year}

\input amssym.def
\input amssym.tex
\font\mybb=msbm10 at 10pt
\def\bb#1{\hbox{\mybb#1}}
\def\bZ {\bb{Z}}

\def\bM {\bb{M}}

\begin{document}


\begin{titlepage}
\vfill
\begin{flushright}
DAMTP-2001-108\\
hep-th/0112077\\
\end{flushright}

\vskip 1cm
\begin{center}
\baselineskip=16pt
{\Large\bf Intersoliton forces in the Wess-Zumino model}
\vskip 0.3cm
{\large {\sl }}
\vskip 10.mm
{\bf ~Ruben Portugues\footnote{e-mail: r.portugues@damtp.cam.ac.uk}
 and ~Paul K. Townsend\footnote{e-mail: p.k.townsend@damtp.cam.ac.uk}}\\
\vskip 1cm
{\small
DAMTP, University of Cambridge, \\
Centre for Mathematical Sciences,
Wilberforce Road, \\
Cambridge CB3 0WA, UK\\
}
\end{center}
\vskip 1cm

\begin{center}
{\bf ABSTRACT}
\end{center}
\begin{quote}
{The spectrum of supersymmetric domain wall solitons of the Wess-Zumino
model is known to be discontinuous across a curve (of marginal stability) in
the moduli space of quartic superpotentials. Here we show how this
phenomenon
can be understood from the behaviour of the long-range inter-soliton force,
which we compute by a method due to Manton.}
\end{quote}

\end{titlepage}

\setcounter{equation}{0}

\section{Introduction}

It is often stated that solitons of supersymmetric field theories that
preserve the same fraction of supersymmetry will exert no force on
each other. When true, this implies the existence of static
supersymmetric multi-soliton solutions that can be interpreted as individual
solitons in static marginal equilibrium due to a cancellation of
attractive and repulsive forces. However, static supersymmetric 
multi-soliton solutions may not exist. This is typically the case
for domain walls; i.e. solitons of (1+1)-dimensional field theories.
Although it is possible to find (1+1)-dimensional models
that admit multi-soliton solutions of the type described 
above \cite{GTT}, these seem to be the exception rather
than the rule. In general, multi-soliton solutions of (1+1)-dimensional
field theories are time-dependent, and this implies the existence of a force
between the constituent solitons.

Consider the case of a field theory for a real scalar field $\phi$ with a
potential having three isolated degenerate minima, at
$\phi_a<\phi_b<\phi_c$.
There exist static soliton solutions interpolating between the adjacent
minima, and these are supersymmetric solutions of the supersymmetric version
of this model, but there is no supersymmetric soliton interpolating
between the {\sl non-adjacent} minima. This is because all
supersymmetric solutions correspond to flows of a first-order equation
and there is no flow connecting $\phi_a$ with $\phi_c$. There must be
{\sl some} solutions that interpolate between the non-adjacent
minima, but they are necessarily time-dependent. These solutions
represent an $ab$-soliton followed by a $bc$-soliton moving under the
influence of the force between them, at least for large separation.
The leading-order term in an asymptotic expansion of the long-range
force must be repulsive because an attractive force would imply the
existence of a bound state of an $ab$-soliton with a $bc$-soliton and hence
the existence of an $ac$-soliton but, as we have just explained, there is
no such soliton. We shall confirm the repulsive nature of the
long-range force, as a special case of a more general result, by
adapting a method introduced by Manton \cite{Manton} to compute the 
long-range attractive force between a soliton and its anti-soliton. 

The situation for {\sl multi-component} scalar field theories is much
more complicated. Here we concentrate on domain walls of the bosonic
sector of the Wess-Zumino model for a single complex
scalar superfield. On reduction to (1+1) dimensions this becomes
a model for a single complex scalar field $z(t,s)$ with
Lagrangian density
\be
\label{lagrangian}
{\cal L} = {1\over2}\left[ |\dot z|^2 - |z'|^2 - |\partial W(z)|^2
\right]\, , 
\ee
where the `superpotential' $W(z)$ is a holomorphic function, and
\be
\dot z \equiv {\partial z\over\partial t}\, ,\qquad z' \equiv
{\partial z\over\partial s}\, ,\qquad
\partial W = {\partial W\over\partial z}\, .
\ee
Critical points of $W$ are degenerate global minima of the potential
and solitons are minimal energy configurations that
interpolate between them. In the supersymmetric context, the   
critical points of $W$ are supersymmetric vacua and the solitons 
interpolating between them preserve 1/2 of the supersymmetry 
\cite{FMVW,AT,CQR}. 

Polynomials provide a simple class of superpotentials. A polynomial
superpotential of order $n$ has $n-1$ critical points, and hence $n-1$
vacua, so in order for such a model to admit multi-soliton configurations we
need $n\ge4$. Here we shall concentrate on the simplest case of a quartic
superpotential,
which may (without loss of generality) be put in the form
\be\label{superP}
W(z)=z^4-\frac{4}{3}\mu z^3-2z^2+4\mu z\,,
\ee
where $\mu$ is a complex constant that parametrizes the space of
physically-distinct quartic superpotentials \cite{AT}. Provided that $\mu
\ne
\pm1$, there are then three (degenerate) supersymmetric vacua, at
\be
(1):\ z=-1, \qquad (2):\ z= 1, \qquad (3):\ z= \mu \ ,
\ee
and hence, potentially, three soliton solutions interpolating between
them. However, whether all three types of soliton actually exist depends on
the
value of $\mu$. For example, if $\mu=0$ then soliton solutions $z(s)$ are
real (because all three vacua lie on the real axis in the $z$-plane) and
can connect only adjacent vacua; there are therefore only two types of
soliton. On the other hand, the choice
\be\label{zthree}
\mu= i \sqrt{3}
\ee
yields a $Z_3$-symmetric model for which the existence of all three
solitons is guaranteed by symmetry given any one of them\footnote{This case
yields an integrable model in (1+1) dimensions \cite{FMVW}.}.

The $\mu$-dependence of the soliton spectrum of the WZ model with
superpotential (\ref{superP}) was analysed in \cite{AT}, where it
was shown that the `2-soliton' to `3-soliton' frontiers in the
$\mu$-plane are the two branches of a curve $\Delta(\mu_1,\mu_2)=0$, where
we have set $\mu=\mu_1 +i\mu_2$ and
\be\label{marginal}
\Delta \equiv 3 -6\mu_1^2 -6\mu_2^2 -\mu_2^4 + 2\mu_1^2\mu_2^2 +3\mu_1^4
\, .
\ee
This curve is shown in Fig. 1; when it is
crossed from inside a `3-soliton' region, one of the
three solitons disappears from the spectrum\footnote{This phenomenom was
rediscovered by other methods in \cite{CV}, where what we call a
WZ model is called a `Landau-Ginzburg' model.}. 
\begin{figure}
\begin{center}
\begin{picture}(0,0)(0,0)
\put (115,130) {\LARGE 3}
\put (115,10) {\LARGE 3}
\put (115,80) {\LARGE 2}
\put (0,80) {\LARGE 2}
\put (210,80) {\LARGE 2}
\end{picture}
\leavevmode
\hbox{%
\epsfysize=2in
\epsfxsize=3in
\epsffile{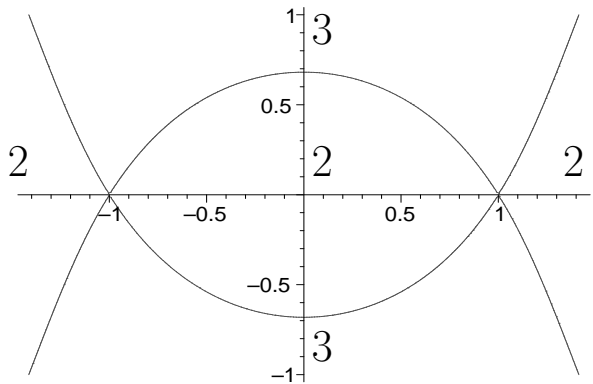}}
\end{center}
\caption[f1]{The curve of marginal stability separating the `2-soliton'
and `3-soliton' regions in the $\mu$-plane, where $\mu=\mu_1+i\mu_2$.}
\end{figure}
The curve $\Delta=0$ is
therefore a simple (and, apparently, the earliest) example of what is now
called a {\sl curve of marginal stability}. The reason for the discontinuity
in the soliton spectrum across this curve was explained in \cite{AT}:
consider
what happens as we start from the $\bZ_3$-symmetric case (\ref{zthree}) and
proceed down the imaginary $\mu$ axis to the origin. As $\mu_2$ decreases,
the
soliton trajectory in the $z$-plane that starts at vacuum 1 ($z=-1$) and
ends
at vacuum 2 ($z=1$) passes increasingly close to vacuum 3 ($z=\mu$). In
other
words, the 12-soliton (connecting vacua 1 and 2) looks increasingly like a
loose bound state of the 13-soliton and 32-soliton. This suggests the
following
picture: near the curve of marginal stability, one of the three solitons can
be
viewed as a bound state of the other two in which the constituents are held
at
a distance that goes to infinity on the curve of marginal stability, thus
causing the bound state soliton to disappear from the spectrum.

The main aim of this paper is to confirm this picture for the
WZ model described above by a determination of the asymptotic intersoliton
force using Manton's method \cite{Manton}. For real
$\mu$ all solitons are real and hence solutions of
a truncated theory involving only a single real scalar field. As
explained above, the two solitons of this theory must repel each
other, at least asymptotically. As we move the $z=\mu$ vacuum away 
from the real axis towards the curve of marginal stability we find 
that this repulsive asymptotic force goes to
zero, changing sign as we cross the curve. So the leading-order force is
attractive on the `3-soliton' side of the curve, as we might expect. We
shall argue, albeit less directly, that the {\it next-to-leading} 
order force is always repulsive. This implies the existence of a bound 
state near the curve of marginal stability (on the `3-soliton' side) 
with a separation of the constituent solitons that diverges on the 
curve. This bound state therefore disappears from the spectrum as the
curve is crossed, in agreement with the results of \cite{AT}. 

A similar result was obtained recently, by different methods, for a
different scalar field theory \cite{RSVV}. We should also note that 
similar results have been obtained, again by different methods,
for monopoles and dyons of (3+1)-dimensional supersymmetric gauge 
theories \cite{RSVV, AN, DGR, RV}. 

We shall begin with a brief review of the solitons of the WZ model with
quartic superpotential. Although they are not known explicitly we will
show that the soliton trajectories in field space can be found {\it
exactly}; this allows some qualitative results of \cite{AT} to be made
quantitative. Next, we show how Manton's computation can be generalized to
yield the asymptotic long-range force between solitons of multi-component
scalar field theories. We then apply this result to the solitons of
the WZ model with quartic superpotential. In particular, we show that
the leading-order force vanishes on the curve of marginal 
stability. We conclude
with a discussion of the next-to-leading order and the implications for
soliton bound states.

\section{WZ solitons for quartic superpotential}

Let $z_i$ $(i=1,2,3)$ be the three critical points of the quartic
superpotential
(\ref{superP}). Given the existence of a soliton connecting vacua $i$ and
$j$,
its topological charge is
\be
T_{ij}=2\left[ W(z_j) -W(z_i)\right]\, .
\ee
These three charges form the three sides of a triangle in the complex
$W$-plane. The triangle inequalty
\be
|T_{ij}| + |T_{jk}| \ge |T_{ik}|
\ee
ensures that any soliton in the spectrum of supersymmetric states must
remain in
it as $\mu$ is varied, unless the triangle degenerates. If this happens,
the
inequality is saturated and the minimum energy configuration with charge
$T_{ik}$ need not be a one-soliton configuration. In fact, since solitons
correspond to straight lines in the $W$-plane\footnote{A recent
demonstration of
this may be found in \cite{MTY}.} only two of the three possible solitons
can exist whenever the three points $W(z_i)$ in the $W$-plane are colinear;
this occurs when $\mu$ is real {\it and} when $\mu$ lies on the curve
$\Delta=0$, with $\Delta$ given by (\ref{marginal}). When $\mu$ is real, the
long-range intersoliton force is repulsive (for reasons explained above)
and,
by continuity, it remains repulsive for $\mu$ in some neighbourhood of the
real axis. Thus, in this region of moduli space only two of the three
possible
soliton states exist and there is therefore no discontinuity in the
soliton spectrum across the real $\mu$ axis. This agrees with the analysis
of
\cite{AT} but it was also shown there, by a study of the qualitative
behaviour
of soliton trajectories in the $z$-plane, that there {\it is} a
discontinuity
across the curve $\Delta=0$. We shall now confirm this by finding the {\it
exact} soliton trajectories.

Supersymmetric solitons solve the first order `BPS' equation
\be
\bar z' = e^{-i\alpha}\partial W(z)\,,
\ee
where $\alpha=\arg (T)$.  Writing $z(x)=u(x)+iv(x)$, this yields the pair
of coupled differential equations
\be\label{bps}
u'  = f(u,v) \, ,\qquad  v' =  g(u,v)
\ee
where, for the quartic superpotential (\ref{superP}),
\bea\label{fandg}
f(u,v) &=&  4\cos\alpha
\left[u^3-3uv^2- u - \mu_1(u^2-v^2-1) + 2uv\mu_2\right] \nn
&& -\ 4\sin\alpha \left[v^3 - 3u^2v +v + 2uv\mu_1 +
(u^2-v^2-1)\mu_2\right]
\nn
g(u,v) &=&  4\cos\alpha \left[v^3 - 3u^2v +v +
2uv\mu_1 + (u^2-v^2-1)\mu_2\right] \nn
&& +\ 4\sin\alpha \left[u^3-3uv^2 -u - (v^2-u^2-1)\mu_1  + 2uv\mu_2\right]
\,.
\eea
These equations determine $u(x)$ and $v(x)$. When $\mu$ is not real we
cannot
find explicit  solutions, but we can find the soliton trajectories in the
$z$-plane. To this end, we observe that (\ref{bps}) implies that $fdv-
gdu=0$.
However,
\be
fdv-gdu =dh
\ee
where
\bea\label{h}
h(u,v) &=& 4\cos\alpha \left[ \left(u^3 -uv^2 -u\right)v - \mu_1v\left(u^2
-{1\over3}v^2 -1\right) +\mu_2u\left(v^2-{1\over3}u^2 +1\right) \right]\nn
&& -\ \sin \alpha \bigg[2v^2 -2u^2 + v^4+u^4 - 6u^2v^2+ 4 \mu_1
u\left(v^2-{1\over3} u^2 +1\right) \nn
&& \qquad \qquad + \ 4\mu_2v \left(u^2 - {1\over3} v^2 - 1\right)\bigg]\,
.
\eea
Soliton trajectories are therefore curves of constant $h$ in the
$z$-plane that pass through two of the critical points. All other curves
of
constant $h$ are `BPS-flows' in that they solve the first-order 
equations (\ref{bps}),
but not with the boundary conditions needed for finite energy.

Consider the 12-soliton interpolating between the vacua at  $z=-1$ and
$z=1$. The curve of constant $h$ that  passes through these points has
\be
h_{12}= {\mu_2\over |\mu|}\, , \qquad \tan\alpha_{12} = {\mu_2\over
\mu_1}\,
,
\ee
The latter relation implies that $\arg T= {\rm actan}\, (\mu_2/\mu_1)$,
which is
consistent with the fact that $T\propto \mu$ for the 12-soliton. The
12-soliton
trajectory is therefore
\bea\label{traj}
0 &=& \mu_2\left[ \left(1-u^2\right)^2 + v^2\left(2+v^2-6u^2\right)\right]
-
4\mu_1 uv\left(u^2-v^2 -1\right) \nn
&&+\ 4|\mu|^2 v\left( u^2-{1\over3}v^2 -1\right) \, .
\eea
In the limit as $\mu\rightarrow\infty$ this trajectory coincides with
the real axis, but for finite $\mu$ it is energetically favourable for
the 12-soliton to partially roll down the potential towards this third
vacuum. A
12-soliton trajectory that passes very close to $z=\mu$ can be viewed as a
bound
state of the 13 and 32 solitons. In the limiting case in which the
trajectory
passes through $z=\mu$ it must either end or begin there, so we then have
an
infinitely separated 13 and 32 soliton but no 12 soliton. Demanding that
(\ref{traj}) pass through $z=\mu$ yields precisely the same condition as
found
in \cite{AT} by requiring colinear topological charges; that is, either
$\mu$ is
real or it lies on the curve $\Delta=0$.

\begin{figure}[!ht]
\begin{center}
\begin{picture}(0,0)(0,0)
\end{picture}
\leavevmode
\hbox{%
\epsfysize=1.8in
\epsfxsize=1.7in
\epsffile{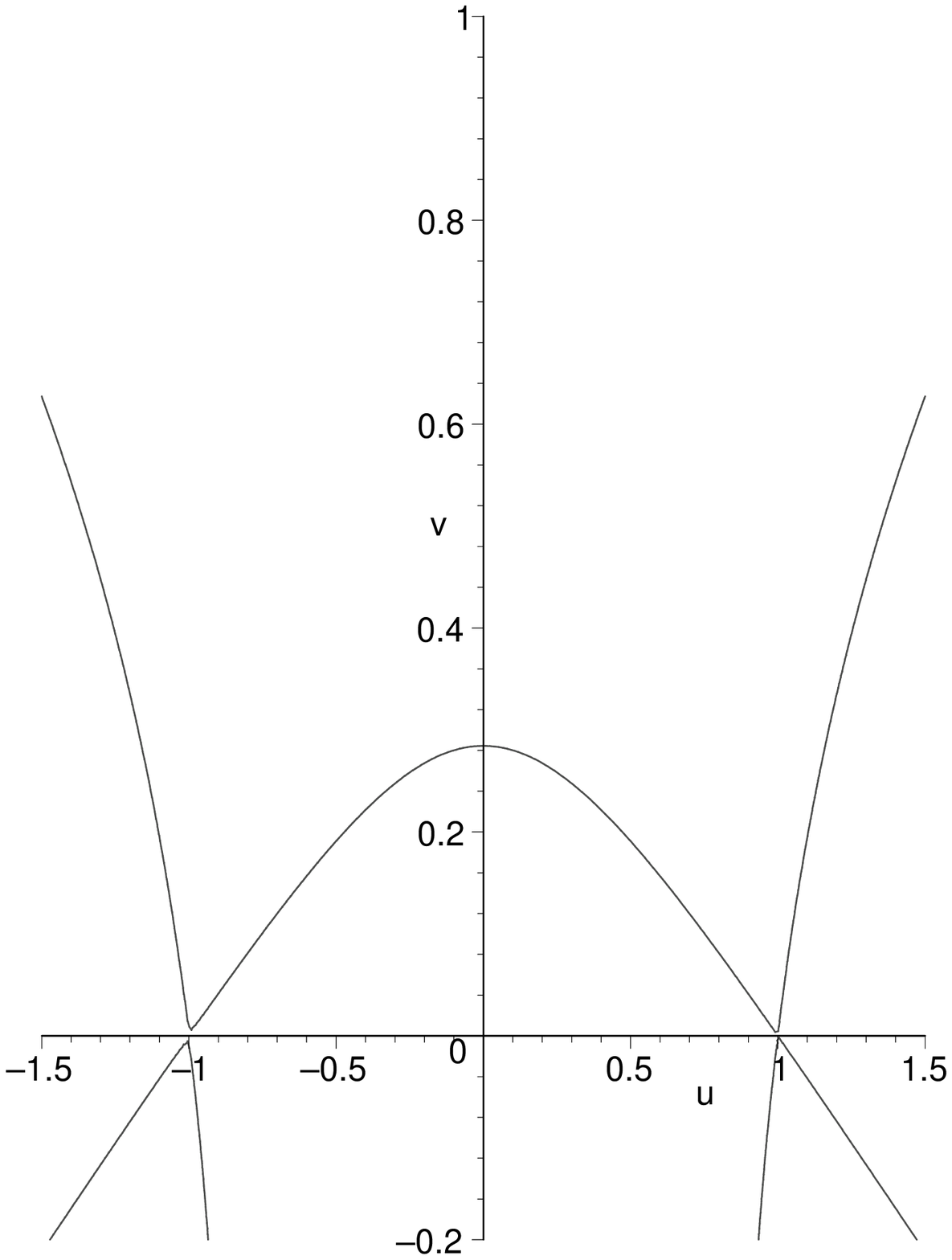}}
\hbox{%
\epsfysize=1.8in
\epsfxsize=1.7in
\epsffile{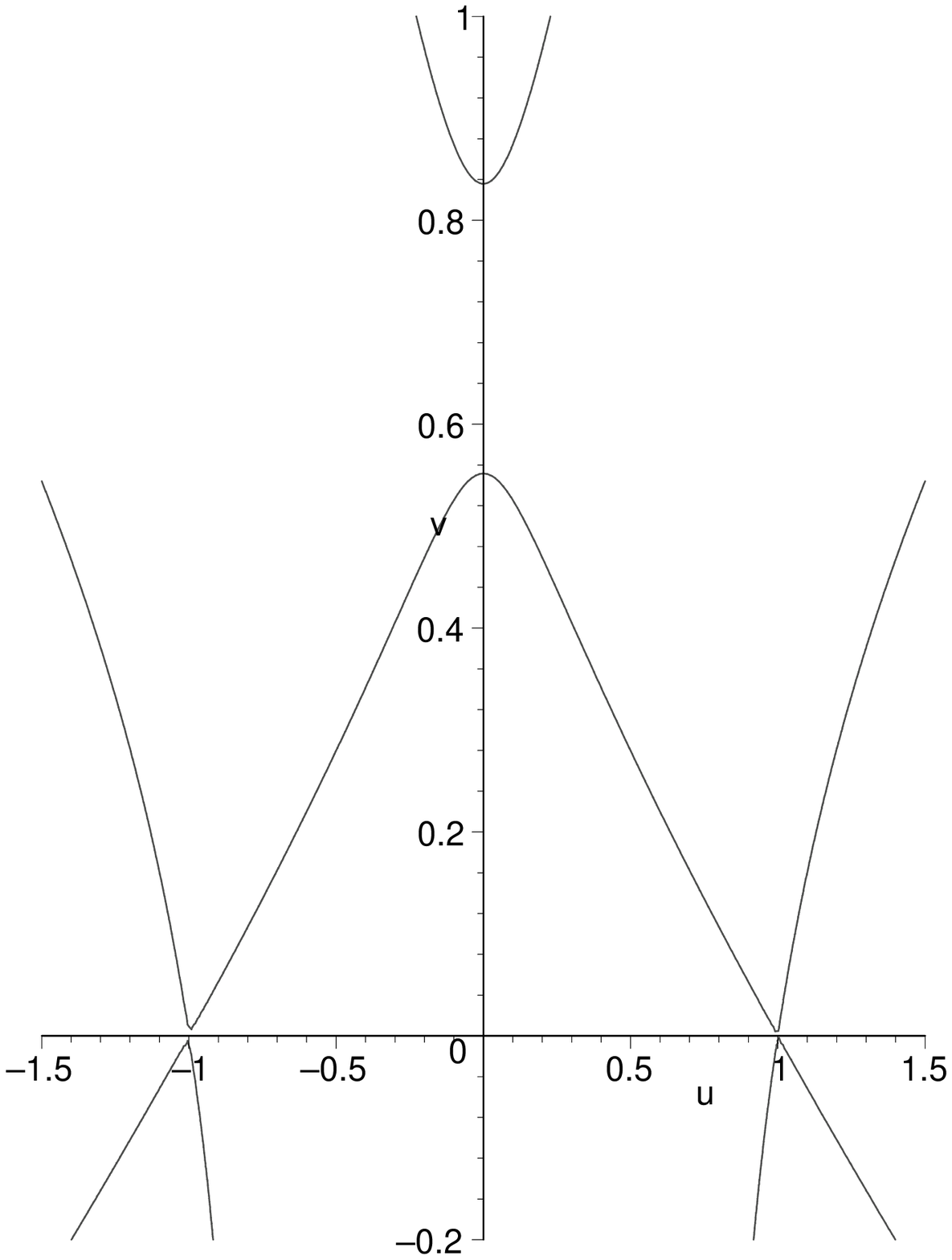}}
\hbox{%
\epsfysize=1.8in
\epsfxsize=1.7in
\epsffile{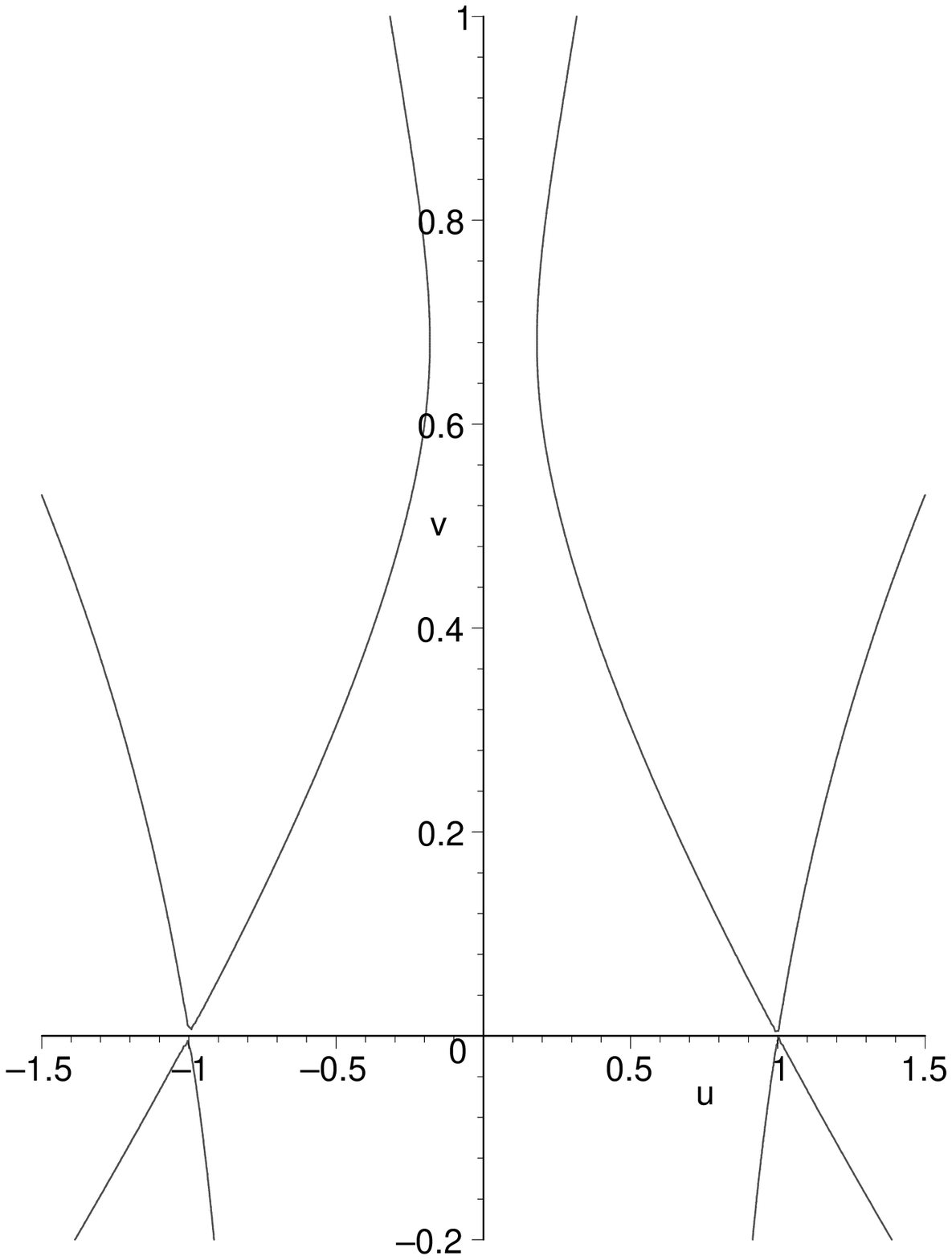}}
\end{center}
\caption[f2]{Plot showing the soliton $S_{12}$ disappearing from
the spectrum as the $z=\mu$ vacuum crosses the curve of marginal
stability in the $z$-plane.  All  plots have $\mu_1=0$ and  
$\mu_2$ is respectively
$1$, $0.7$ and $0.65$.  The critical value for which $\Delta=0$ if
$\mu_1=0$ is $\mu_2= \pm \sqrt{2\sqrt{3}-3} \approx \pm 0.681$.  }
\end{figure}

The other two soliton trajectories can be found similarly. In each
case the soliton trajectory coincides with the union of the other two
when $\mu_2\Delta=0$, although it disappears from the spectrum only on
crossing the curve $\Delta=0$. It should be noted that
the relevant segments of the curve
$\Delta=0$ differ in all three cases. The above case, in which the
12-soliton is the one that disappears from the spectrum as we go from
a `3-soliton' to a `2-soliton' region, corresponds to the segments with
$\mu_1^2<1$, whereas the other two cases correspond to segments with
$\mu_1^2>1$. We skip the details but note for future use that
\be\label{alphas}
\tan\alpha_{13} = {\mu_2 \over \mu_1 + \Delta_+}\, ,\qquad
\tan\alpha_{23} = {\mu_2 \over \mu_1 + \Delta_-}\, ,
\ee
where
\be
\Delta_\pm = \pm {\Delta \over 4\left[ 3\mu_1 +
\mu_1\mu_2^2 -\mu_1^3 \pm 2\right]}\, .
\ee
It follows that $\alpha_{13}=\alpha_{32}$ when $\Delta=0$, as expected.

\section{The leading-order long-range force}

We now plan to obtain a formula for the asymptotic force between two
solitons of
a (1+1)-dimensional field theory for a multi-component real scalar field
$\phi(t,s)$ with
Lagrangian density
\be
{\cal L} =  {1\over2} |\dot\phi|^2 - {1\over2}|\phi'|^2 -V\, .
\ee
This of course includes (\ref{lagrangian}) as a special case of a 
two-component real scalar field Lagrangian density. 
We assume that $V$ has multiple degenerate vacua at $\phi=\phi_i$
$(i=1,2,\dots)$, and that some vacua are connected by static soliton
solutions.
Let $\phi_{ij}(s)$ be a soliton connecting the $i$th vacuum to the $j$th
vacuum.
Near the $i$th vacuum we have
\be
V \approx {1\over2}m_i^2 |\phi -\phi_i|^2\, .
\ee
The asymptotic form of the $ij$-soliton solution near the $j$th vacuum (as
$s\rightarrow\infty$) is therefore
\be
z_{ij}(s) - z_j \sim - m_j^{-1} t_{ij} e^{-m_js}\, ,
\ee
where $t_{ij}$ is tangent to the $ij$-soliton trajectory in field space as
it
approaches the $j$th vacuum. Similarly, the asymptotic form of the
$ij$-soliton
near
the $i$th vacuum (as $s\rightarrow -\infty$) is
\be
z_{ij}(s) -z_i \sim m_i^{-1} t_{ji} e^{m_is}\, .
\ee

Now consider a static configuration of an $ik$-soliton at the origin, $s=0$,
separated by a distance $L$ from a $kj$-soliton at $s=L$.  A configuration
that
fits this description is
\be\label{ansatz}
\phi(s) = \phi_{ik}(s) + \phi_{kj}(s-L) - \phi_k\, .
\ee
We should not expect this to be an exact solution of the field equations
but
it
will be an approximate solution for large $L$ provided that there is a $b$
in
the range $0\ll b\ll L$ such that both individual static soliton solutions
$\phi_{ik}(s)$ and $\phi_{kj}(s)$ are nearly equal to their common vacuum
value
$\phi_k$ near $s=b$. We shall write (\ref{ansatz}) as
\be\label{ansatz2}
\phi(s) = \phi_{ik}(s) + \varphi(s)
\ee
where
\be
\varphi(s) = \phi_{kj}(s-L) - \phi_k
\ee
and assume that $\varphi(s)$ is a small perturbation to the static
solution
$\phi_{ik}(s)$ in the region $s\le b$. Since, for $s\approx b$,
\be\label{asymp2}
\phi_{ik}(s) - \phi_k \sim - m_k^{-1} t_{ik} e^{-m_ks}\, ,
\qquad \varphi(s) \sim m_k^{-1} t_{jk}e^{m_k(s-L)}\, ,
\ee
this means that we must have $2b\ll L$, so we are now assuming that
\be 
0 \ll 2b\ll L\, .
\ee

The force exerted on the ik-soliton by the kj-soliton can now be found as
follows \cite{Manton}: the momentum of the ik-soliton is approximately
given
by\footnote{The sign, which agrees with \cite{Manton}, is chosen such that
a
negative force corresponds to an attractive intersoliton potential.}
\be
P = -\int_{-\infty}^b \dot\phi \cdot \phi'\, ds\, .
\ee
The force on it is therefore
\be\label{force}
F \equiv \dot P = \left[ -{1\over2}|\dot \phi|^2 - {1\over2}|\phi'|^2 +
V \right]_{-\infty}^b \, ,
\ee
which follows on use of the field equation
\be\label{fe}
\ddot \phi = \phi'' - {\partial V\over \partial \phi}\, .
\ee
Using the ansatz (\ref{ansatz2})
and properties of $\phi_{ik}$, notably
\be
|\phi_{ik}'|^2 = 2V(\phi_{ik})\, ,
\ee
we find that
\be\label{force2}
F= - \left[\phi_{ik}' \cdot \varphi' - \phi_{ik}'' \cdot \varphi \,
\right]_{-\infty}^b + {\cal O}(\varphi^2)\, .
\ee
This may be evaluated using the asymptotic forms (\ref{asymp2}) of
$\phi_{ik}$
and $\varphi$. The result is
\be
F= - 2 t_{ik}\cdot t_{jk}\, e^{-m_kL}\, .
\ee

This generalizes the result of Manton to multi-component scalar field
theories. 
For a single-component theory the tangent vectors are necessarily either
parallel or antiparallel, so the force is either attractive or repulsive,
never
zero. For example, for a soliton-antisoliton pair we have $i=j$ and hence
the
attractive force \cite{Manton}.
\be
F= -2 t_{ik}^2 e^{-m_kL} \, .
\ee

\section{Application to the WZ model}

The above results are applicable to the WZ model because this has degenerate
vacua
at critical points $z=z_i$ of the superpotential $W$, with
\be
m_i^2 = |W'(z_i)|^2\, .
\ee
We shall consider again the case of a quartic superpotential with critical
points at $z=-1,1,\mu$ (with $\mu=\mu_1+ i\mu_2$), and choose
\be
\mu_1^2 <1\, ,
\ee
so that the 12-soliton is the one expected to appear as a bound state (of
a 13-soliton and 32-soliton) near the curve of marginal stability.
Consider a field configuration in which a 13-soliton is at $s=0$
and a 32-soliton at $s=L$, for large $L$. In the region between
these
solitons, and far from both, we may linearize the first-order equations
(\ref{bps}) about the $z=\mu$ vacuum to get
\be\label{linear}
\pmatrix{u'\cr v'} = \bM \pmatrix{u-\mu_1\cr v-\mu_2}
\ee
where the matrix $\bM$ is
\be
\bM = 4\pmatrix{\mu_1^2-\mu_2^2 -1  &2\mu_1\mu_2\cr 2\mu_1\mu_2&
\mu_1^2-\mu_2^2 -1}\pmatrix{\cos\alpha & \sin\alpha \cr -\sin\alpha
&\cos\alpha}\, .
\ee
This matrix has eigenvalues $\pm m_3$, where
\be
m_3 = 4\sqrt{1+ |\mu|^4}\, .
\ee
The corresponding eigenvectors are
\be
e_\pm(\alpha) = {1\over \sqrt {2}\left[1\mp
\cos(\theta+\alpha)\right]^{1\over2}}
\left(\sin(\alpha +\theta),  -\cos(\alpha +\theta) \pm 1\right)
\ee
where
\be\label{tantheta}
\tan\theta = {2\mu_1\mu_2\over 1+\mu_2^2 -\mu_1^2}\, .
\ee
These eigenvectors are tangents to the separatrix flows at $z=\mu$. Note
that they are {\it orthogonal}.

The generic solution of (\ref{linear}) has both $e^{-m_3s}$ and $e^{m_3s}$
terms,  but the asymptotic 13-soliton solution $z_{13}$ has only the
$e^{-m_3s}$
exponential term. Thus, we should take
\be
t_{13}\propto e_-(\alpha_{13}) \, .
\ee
Similarly, the asymptotic 32-soliton solution, near the 3-vacuum, has only
the
$e^{m_3s}$ factor, so we should take
\be
t_{23} \propto e_+(\alpha_{32})\, .
\ee
The sign of the constant of proportionality should be the same in both
cases,
so this yields the force
\be
F \propto - \sin\left[{1\over2}(\alpha_{13}-\alpha_{32})\right]
\sin\left[{1\over2}(\alpha_{13} +\alpha_{32}) +\theta\right]
+ \sin^2\left[{1\over2}(\alpha_{13}-\alpha_{32})\right]\,
\ee
for some positive constant of proportionality.

Consider first the case for which $\mu_2=0$, so that both
soliton solutions correspond to BPS-flows along the real axis in the
$z$-plane
from $z=-1$ to $z=1$ (since $\mu_1^2<1$ by hypothesis). 
From (\ref{alphas}) we then see that
\be
\alpha_{13} = 0,\pi\, .
\ee
Similarly for $\alpha_{32}$ but $\alpha_{32} = \alpha_{13} + \pi$.
To see why, note that a BPS flow just above the real axis will approach
the vacuum at $z=\mu$ anti-parallel to the real axis ($e_-$) and then
leave
it
parallel to the imaginary axis ($-e_+$). To arrange for this flow to reach
$z=1$
(or close to it) we must rotate $-e_+$ back to $e_-$. This is achieved by
a
shift of the angle by $\pi$ since
\be
e_\mp(\alpha+\pi) = -e_\pm(\alpha)\, .
\ee
Thus, in this case,
\be
\alpha_{32} = \alpha_{13} +\pi\, ,
\ee
and this leads to $F \propto (1-\sin\theta)$. Since $\sin\theta =0$ for
real $\mu$ we thus deduce that $F\propto 1$ with positive constant of
proportionality; that is, $F>0$, so the force is repulsive, as
expected. 

We now turn to those cases for which $\mu$ is near the curve of marginal
stability. In this case we see from (\ref{alphas}) that
\be\label{angdiff}
\left(\alpha_{13}-\alpha_{23}\right) = {-\mu_2\Delta \over 4\mu_1^2 (1
-\mu_1^2)^3} + {\cal O}\left(\Delta^2\right) \, .
\ee
It follows, after some calculation, that
\be
\label{nearmforce}
F \propto {\mu_2^2 \left(1+ |\mu|^2\right)
\left(1+\mu_2^2-\mu_1^2\right) \Delta \over (1-\mu_1^2)^3
\sqrt{\mu_1^2 (1- |\mu|^2)^2 +
\mu_2^2(1+|\mu|^2)^2\left(1+\mu_2^2-\mu_1^2\right)^2} }
+ {\cal O}\left(\Delta^2\right)\,.
\ee
Since $\mu_1^2<1$, all factors other than $\Delta$ are positive. We
thus have
\be\label{forceL}
F(L) \sim c^2\left[\Delta + {\cal O}\left(\Delta^2\right)
\right] e^{-m_3L}\, ,
\ee
for some non-zero constant $c$. The leading-order
asymptotic force is therefore repulsive when $\Delta>0$, which
corresponds to a point inside the `2-soliton' region. It is
attractive when $\Delta<0$, which corresponds to a point inside
a `3-soliton' region. On the curve of marginal stability the
leading-order asymptotic force is zero.

\section{Soliton bound states}

So far, we have analysed the behaviour of solitons in the WZ model with
quartic superpotential near the curve of marginal stability in the moduli
space of such superpotentials. On one side of this curve there are only two
types of soliton and the long-range asymptotic force between these two
solitons is repulsive. We have shown, however, that this repulsive asymptotic 
force vanishes on the curve of marginal stability, and becomes
attractive after it is crossed.

An interesting question that this result raises is whether the long-range
intersoliton force vanishes on the curve of marginal stability only to
leading order in an asymptotic expansion (in powers of $e^{-m_3L}$)
or to all orders (assuming the existence of this expansion). If the
force were to vanish {\it exactly} on the curve of marginal stability
then we would expect to be able to find a one-parameter family of 
12-soliton solutions corresponding to
a 13 and 32 soliton at arbitrary separations. However, there are no
such solutions because the 12-soliton trajectory necessarily 
coincides, on (the appropriate segment of) the curve of 
marginal stability, with the union of the 13 and 32-soliton 
trajectories. This fact suggests that it is {\it
only} the leading-order asymptotic force that vanishes on the curve of
marginal stability, and that the next-to-leading order term
is non-zero. Given that it {\it is} non-zero it must 
be repulsive because an attractive force would imply a bound 
state `third soliton' on the `2-soliton' side of the curve of 
marginal stability. Near $\Delta=0$ we thus expect an
asymptotic expansion for the intersoliton force of the form
\be\label{asympF}
F(L) \sim c^2\Delta e^{-m_3L} + \gamma^2 e^{-2m_3L} + \dots
\ee
for some non-zero constant $\gamma$. When $\Delta <0$ this force vanishes for
\be
e^{-m_3L} \approx \left(c^2/\gamma^2\right)|\Delta|\, .
\ee
Since $\Delta$ is small the neglect of the higher-order terms in
(\ref{asympF}) is justified, and there is thus a minimum of the
inter-soliton
potential at $L\sim (1/m_3)\log (1/|\Delta|)$. This diverges on the curve of
marginal stability, as we know must happen from the analysis of the soliton
trajectories, and this explains the discontinuity of the spectrum on
crossing this curve. 
As mentioned earlier, this result agrees qualitatively with a similar result
obtained for a different scalar field model by different methods
\cite{RSVV}, as well as with results for other types of 
supersymmetric soliton.

\vskip 1cm
\noindent
{\bf Acknowledgements}: We thank Jerome Gauntlett and Nick Manton for
very helpful discussions.  R.P. thanks Trinity College Cambridge for
financial support.

\end{document}